\documentclass[11pt]{article}
\usepackage{amsmath, amssymb, amsthm,fullpage, epic, eepic}
\usepackage[all]{xypic}
\newtheorem{thm}[equation]{Theorem}
\newtheorem{prop}[equation]{Proposition}

\newtheorem{lemma}[equation]{Lemma}

\theoremstyle{definition}
\newtheorem{defn}[equation]{Definition}

\theoremstyle{remark}

\newtheorem{ntn}[equation]{Notation}

\newtheorem{rem}[equation]{Remark}

\makeatletter
\renewcommand{\subsection}{\@startsection{subsection}{2}{0pt}{-3ex
plus -1ex minus -0.2ex}{-2mm plus -0pt minus
-2pt}{\normalfont\bfseries}} 
\renewcommand{\subsubsection}{\@startsection{subsubsection}{2}{0pt}{-3ex
plus -1ex minus -0.2ex}{-2mm plus -0pt minus
-2pt}{\normalfont\bfseries}} \makeatother

\numberwithin{equation}{subsection}

\newdir^{ (}{{}*!/-5pt/@^{(}}

\newcommand{\tra}{\mathop{\rightarrow}}

\newcommand{\iso}{{\;\stackrel{_\sim}{\to}\;}}

\newcommand{\beq}{\begin{equation}\label}
\newcommand{\eeq}{\end{equation}}

\newcommand{\into}{\hookrightarrow}

\newcommand{\Hom}{\operatorname{Hom}}

\newcommand{\op}{\text{op}}
\newcommand{\g}{\mathfrak{g}}
\newcommand{\T}{\mathcal{T}}

\newcommand{\ord}{\operatorname{ord}}

\def\R{\mathbb{R}}

\def\k{\mathbf{k}}

\def\o{\otimes}

\def\dq{{\overline{Q}}}

\def\Id{\mathrm{Id}}

\def\Syns{{\operatorname{Sym}}}
\def\Sym{{\operatorname{Sym\ }}}
\def\Z{{\mathbb Z}}

\def\1{\mathbf{1}}

\begin{document}
\title{The necklace Lie coalgebra and renormalization algebras}
\author{Wee Liang Gan and Travis Schedler} 
\maketitle
\begin{abstract}
  We give a natural monomorphism from the necklace Lie coalgebra,
  defined for any quiver, to Connes and Kreimer's Lie coalgebra of
  trees, and extend this to a map from a certain quiver-theoretic Hopf
  algebra to Connes and Kreimer's renormalization Hopf algebra, as
  well as to pre-Lie versions.  These results are direct analogues of
  Turaev's results in 2004, by replacing algebras of loops on surfaces
  with algebras of paths on quivers. We also factor the morphism
  through an algebra of chord diagrams and explain the geometric
  version. We then explain how all of the Hopf algebras are uniquely
  determined by the pre-Lie structures, and discuss noncommutative
  versions of the Hopf algebras.
\end{abstract}

\section{Introduction}
Motivated by an attempt to understand the moduli space of flat
connections on a vector bundle over a surface, Goldman constructed in
\cite{Go} a Lie bracket on the free vector space spanned by homotopy
classes of (basepoint-free) loops on a surface, together with a Lie
homomorphism (by taking trace of holonomy) to the Poisson algebra of
functions on the aforementioned moduli space.

In \cite{Tu1}, Turaev discovered that one may define, in a similar
way, a cobracket on the aforementioned space of loops, which is
compatible with Goldman's bracket and yields a Lie bialgebra. He also
constructed a quantization of this Lie bialgebra in terms of link diagrams
on the surface.

This left open the question to find an interpretation of the Lie
coalgebra.  In \cite{Tu2}, Turaev discovered a relation between his
Lie coalgebra and Connes and Kreimer's renormalization algebras
\cite{K,CK}, which form part of the algebraic foundations of
perturbative quantum field theory.  Specifically, he found a
homomorphism from an up-to-isotopy, pointed version of his Lie
coalgebra of loops to Connes and Kreimer's Lie coalgebra of trees, and
constructed from this a commutative Hopf algebra on the loop side\footnote{This Hopf algebra differs significantly from the Hopf algebra mentioned in the previous paragraph: aside from the previous one referring to homotopy classes rather than isotopy classes, the Hopf algebra here is commutative, unlike the one of the previous paragraph.}
mapping to Connes and Kreimer's renormalization Hopf algebra.  This
allows one to interpret the combinatorics of loops and Turaev's
Lie coalgebra as Feynman diagrams connected to renormalization.

Turaev found that additional structure from the
construction is preserved, and accordingly generalized Connes and
Kreimer's algebras to include the extra data. This attached to the
trees the following: (1) a homotopy class of loops assigned to each vertex;
(2) orientations on the edges; and (3) a ribbon graph structure.

An essential step of the above construction is the observation that,
when one has a basepoint, Turaev's Lie coalgebra actually arises from
a more fundamental pre-Lie coalgebra structure.  In the basepoint-free
case, Turaev constructed an ``oriented trees'' version of Connes and
Kreimer's Lie coalgebra (which does \textbf{not} come from a pre-Lie
coalgebra), so that one is still equipped with a homomorphism.

There is a known analogue for quivers of the Goldman/Turaev Lie
bialgebra of loops, called the necklace Lie bialgebra.  Here, the Lie
algebra was discovered for much the same reason as the Goldman
algebra: because of its representation into the Poisson algebra of
functions on the corresponding quiver variety \cite{Gi} (the Lie
algebra was independently discovered in \cite{BLB}). In \cite{S},
the second author constructed the cobracket and quantized the
resulting Lie bialgebra, following in analogy with \cite{Tu1}. It is
thus natural to ask for an interpretation of the cobracket in terms of
representations, and in particular, if all of the above results
from \cite{Tu2} can be generalized to the quiver setting.

In this note, we answer this question affirmatively, and present
quiver analogues of the results of \cite{Tu2}.  We define a pre-Lie
coalgebra on the quiver side in the rooted (basepointed) case.  We
also define a commutative ``renormalization'' Hopf algebra associated
to any quiver.\footnote{As in Turaev's case, this Hopf algebra is
  quite different from the quantized necklace algebra of \cite{S}: the
  latter is noncommutative, unlike the former; and the former involves
  paths which are not mod commutators, unlike the latter.}
Then, our main theorem (Theorem \ref{homt1}) is the construction of a
natural monomorphism from these quiver algebras and the (oriented)
necklace Lie coalgebra to the Connes and Kreimer algebras, with
additional quiver-theoretic structure we define.  We explain how the
pre-Lie and Hopf algebra structures are essentially equivalent, in all
cases (necklaces, trees (Connes-Kreimer), and loops (Turaev)), using a
general result about pre-Lie algebras \cite{GuOu}.

Unlike in Turaev's case, for quivers, there is no distinction between
``up to isotopy'' and ``up to homotopy,'' so there is only one Lie
coalgebra to consider in each of the oriented/rooted cases, which is
the one that is compatible with the necklace bracket in the oriented
case. In contrast, the isotopy coalgebra from \cite{Tu2} is
\textbf{not} compatible with the Goldman Lie bracket (compatibility
requires passing to homotopy classes).

For quivers, we find that introducing a basepoint is the same as
cutting a necklace and considering algebras of paths, so our algebras
in the rooted case are actually algebras of paths in the quiver.  The
additional structure attached to trees in our setting replaces the
loops (with basepoint) attached to vertices with cyclic paths (paths)
in the quiver.  The rest of the structure---the ribbon graph structure and
orientation of edges---is unchanged.

Furthermore, we discover that the monomorphisms factor through a
``chord algebra'' we define, which essentially is the span of
necklaces with certain chord diagrams (Theorem \ref{homt2}).  This
clarifies the construction and the reason that it exists.  We also
briefly describe the corresponding geometric object, in the setting of
\cite{Tu2}, where the chord diagrams become geometric chord diagrams
\cite{AMR} (the chord algebras were not mentioned in \cite{Tu2}).

Finally, we briefly give a noncommutative version of the Hopf algebra,
analogous to Section 8.5 of \cite{Tu2}, which maps to Foissy's
noncommutative algebra of labeled rooted trees \cite{F}.  In particular, this
shows that paths and loops have a noncommutative (=ordering) structure
which does not exist for rooted trees without labels.  The
noncommutative Hopf algebras, unlike their commutative counterparts,
are \textbf{not} determined by the pre-Lie structure alone (which
essentially forgets about the ``labelings'' on the associated trees).

The organization of the paper is as follows: in Section 2, we briefly
recall the necessary definitions from \cite{K, CK,Tu2} (for the tree
side), and from \cite{S} (for the quiver side).  Then, in Section 3,
we define the new quiver-theoretic Hopf algebra and pre-Lie coalgebra,
state our main result, and generalize it through chord algebras. We
also explain the equivalence of pre-Lie coalgebras and commutative
Hopf algebras of a special form, and give the noncommutative version
of the constructions in this paper.  Finally, in Section 4, we provide
the postponed proofs (e.g., of the main theorem).
\subsection{Acknowledgements}
We are grateful to Victor Ginzburg for connecting the authors and for
some useful comments. We thank Muriel Livernet for helpful comments and
references.
The first author was partially supported by NSF grant DMS-0726154. The
second author was partially supported by an NSF GRF.
\subsection{Notation}
\begin{ntn} Throughout, $\k$ denotes a fixed commutative ring with unit.
\end{ntn}
\begin{ntn} \label{tauntn} For any permutation $\sigma \in
  \Sigma_{n}$,
  we define
  $\tau_\sigma: V_1 \o V_2 \o \cdots \o V_n \rightarrow
  V_{\sigma^{-1}(1)} \o V_{\sigma^{-1}(2)} \o \cdots \o
  V_{\sigma^{-1}(n)}$ as the   permutation of components corresponding to $\sigma$.  
\end{ntn}
\begin{ntn} Our permutations use cycle notation. Thus, $(123)$ denotes the
permutation $1 \mapsto 2 \mapsto 3 \mapsto 1$.
\end{ntn}

\section{Tree algebras}
In this section, we recall the needed constructions of \cite{Tu2, K, CK}.
\subsection{Algebras of trees}
We recall the Lie coalgebra and Hopf algebra of rooted trees from
\cite{K, CK}, following \cite{Tu2}, where the former is generalized to a
pre-Lie coalgebra, and to the setting of oriented trees.  

\subsubsection{Pre-Lie (co)algebras}
We recall first the definition of pre-Lie algebras (independently
discovered by \cite{Ge} and \cite{Vi}) and their dual, pre-Lie
coalgebras (following \cite{Tu2}).  Note that there are a wide variety
of important pre-Lie algebras, including the Hochschild cochain complex
of an algebra, vector fields, and the examples in this paper.

A (left) pre-Lie algebra over $\k$ is a $\k$-module with a $\k$-bilinear 
product $\star$ satisfying
\begin{equation}
(x \star y) \star z - x \star (y \star z) = (y \star x) \star z - y
\star (x \star z).
\end{equation}
If $\star$ is a pre-Lie multiplication, then
$[x,y] := x \star y - y \star x$ must be a Lie bracket.

To obtain the definition of pre-Lie coalgebra, we dualize in the sense
of determining what structure exists on $\mathfrak{g}^*$ if
$\mathfrak{g}$ is a pre-Lie algebra.  Precisely, a (left) pre-Lie
coalgebra, $\mathfrak{g}$, over $\k$ is a $\k$-module with a
$\k$-linear map $\delta_0: \g \rightarrow \g \otimes \g$ satisfying
\begin{equation}
(\Id - \tau_{(12)}) (\delta_0 \o 1 - 1 \o \delta_0) \delta_0 = 0 \in \Hom_\k(\g, \g \o \g \o \g).
\end{equation}
If $\delta_0$ is a pre-Lie comultiplication, then
$\delta := \delta_0 - \tau_{(12)} \delta_0$ must be a Lie cobracket.

\subsubsection{The pre-Lie coalgebra of rooted trees and Lie coalgebra
of oriented trees}
A rooted tree $T$ is a collection of vertices $V(T)$ and edges $E(T)$,
and a map $E(T) \rightarrow V(T)^{(2)}$ from edges to unordered pairs
of vertices, such that the resulting graph is connected and has no
cycles (or loops), together with a distinguished vertex, called the
root.

An oriented tree is the same but without the distinguished vertex, and whose
edge map is actually a map $E(T) \rightarrow V(T)^2$, from edges to ordered
pairs of vertices.  

Let $\T_{rt}$ be the free $\k$-module with basis given by the
isomorphism classes of rooted trees.  In \cite{CK} (\cite{Tu2} for
the ``pre-''), the following pre-Lie comultiplication $\rho$ is
defined on $\T_{rt}$:
\begin{equation} \label{comroot}
\rho(T) = \sum_{e \in E(T)} T_e^1 \o T_e^2,
\end{equation}
where $T_e^1, T_e^2$ are the trees obtained by deleting the edge $e$, and
$T_e^2$ is the tree which contains the root. The root of $T_e^1$ is the vertex
which was incident to $e$, and the root of $T_e^2$ is the root of $T$.
\begin{prop}\cite{Tu2} The map $\rho$ is a pre-Lie comultiplication.
\end{prop}
Thus, one deduces that the skew-symmetrization,
$\rho^{ss} := \rho - \tau_{(12)} \rho$, is a Lie cobracket, which was
already discovered in \cite{CK} (and motivated the above result).

In \cite{Tu2}, a version for oriented trees is also given. Let
$\T_{or}$ be the free $\k$-module with basis given by the isomorphism
classes of oriented trees.  Then, a Lie cobracket
$\rho^{ss}_{O}: \T_{or} \rightarrow \T_{or} \o \T_{or}$ is defined by
the same formula as the skew-symmetrization of \eqref{comroot}, except
letting $T_e^2$ be the subtree that $e$ points to, and $T_e^1$ be the
subtree that $e$ points away from (note that there is no pre-Lie
comultiplication $\rho_O$).

Finally, we will need the generalization given in \cite{Tu2}: Let
$RTrees, OTrees$ be the categories whose objects are rooted and
oriented trees, respectively, and whose morphisms are embeddings of
trees (maps of rooted or oriented trees, that preserve incidence and
are injective on vertices and edges; the root must get sent to the
vertex of the image subtree which is closest to the root of the whole
tree).  Then, we have
\begin{defn}
For any contravariant
functor $\Phi: RTrees \rightarrow Sets$ (called a \textbf{rooted
  tree-structure}),
we let $\mathcal{T}_{rt}(\Phi)$ be the free $\k$-module spanned by
isomorphism classes of pairs $(T,s)$ where $T$ is a rooted tree and
$s \in \Phi(T)$. Here, an isomorphism of pairs $(T,s) \iso (T',s')$ is
an isomorphism of trees $T \iso T'$ whose pullback carries $s'$ to
$s$.  In the oriented case, one similarly defines $\T_{or}(\Phi)$.
\end{defn}
\begin{prop}\cite{Tu2} For any rooted tree-structure $\Phi$,
the following formula defines a pre-Lie comultiplication
$\rho$ on $\T_{rt}(\Phi)$:
\begin{equation}
\rho(T,s) = \sum_{e \in E(T)} (T_e^1,s|_{T_e^1}) \o
(T_e^2, s|_{T_e^2}),
\end{equation}
where $T_e^1 \sqcup T_e^2 = T \setminus \{e\}$ and $T_e^2$ contains
the root.  Similarly, the skew-symmetrization of this defines a Lie
coalgebra in the case of oriented trees, where now
$T_e^2$ is the tree that $e$ points to.
\end{prop}

\subsubsection{The Hopf algebra on rooted trees}
We briefly recall Connes and Kreimer's Hopf algebra on rooted trees
\cite{K, CK}, as formulated with tree structures in \cite{Tu2}.

Let $\Syns(\T_{rt})$ be the symmetric algebra on $\T_{rt}$ (polynomials in rooted
trees).
\begin{defn}\cite{CK} A \textbf{cut} $H$ of a rooted tree $T$ 
  with root $rt \in V(T)$ is a subset $H \subset E(T)$ of edges. It is
  a \textbf{simple cut} if each connected component of $T \setminus
  \{rt\}$ contains at most one edge in $H$. (The empty cut $H = \emptyset$ is
  included, and is simple.)
\end{defn}
\begin{defn}\cite{CK} For any simple cut $H$
  of $T$, let $\{T_{H,e}: e \in H\} \cup \{T_{H,0}\}$ be the set of
  connected components of $T \setminus H$, where $T_{H,0}$ is the
  component containing the root $rt$, and $T_{H,e}$ is the other
  connected component which was adjacent to $e$ as a subset of $T$.
\end{defn}
\begin{defn}\cite{CK,Tu2}
For any rooted tree-structure $\Phi: RTrees^{\op} \rightarrow Sets$, 
define a map $\Delta: \T_{rt}(\Phi) \rightarrow \Sym \T_{rt}(\Phi) \o \Sym \T_{rt}(\Phi)$ 
by the formula
\begin{gather}
  l_H(T,s) := \prod_{e \in H} (T_{H,e}, s|_{T_{H,e}}) \in \Sym \T_{rt}(\Phi), \quad
  r_H(T,s) = (T_{H,0}, s|_{T_{H,0}}), \\
  \Delta(T,s) =  (T,s) \o 1 + \sum_{\text{simple cuts } H} l_H(T,s) \o r_H(T,s).
\end{gather}
Here, by definition, $l_{\emptyset}(T,s) = 1$ and $r_{\emptyset}(T,s) = (T,s)$,
for all $T,s$.
\end{defn}
\begin{prop}\cite{CK,Tu2} This defines a commutative Hopf algebra structure 
  on $\Sym \T_{rt}(\Phi)$ for any rooted tree-structure $\Phi$, with
  counit $\epsilon(X) = 0$ for any $X = (T,s)$.
\end{prop}
Note that comultiplication has the form (for $X = (T,s)$)
\begin{equation}
\Delta(X) = 1 \o X + X \o 1 + \Delta'(X),
\end{equation}
where $\Delta'$ is the projection of $\Delta$ away from
$(1 \o \Sym \T_{rt}(\Phi)) \oplus (\Sym \T_{rt}(\Phi) \o 1)$ (that is,
$\Delta' = (1 - \eta \circ \epsilon)^{\o 2}\Delta$, with
$\eta$ the unit map).

Furthermore, using the natural grading by total number of edges in the trees,
$\Delta'(X)$ has strictly lower degree than $X$ if $X$ is a tree.
Thus,
one may easily verify (cf.~\cite{S}, \S 3.9) that the following
general formula for the antipode $S$ holds (with $X = (T,s)$):
\begin{equation} \label{afla}
S(X) = -X + \sum_{n \geq 1} (-1)^{n+1} \mu^n \circ (\Delta')^n(X),
\end{equation}
(where $\mu^n:
\Sym \T_{rt}(\Phi)^{\o (n+1)} \rightarrow \Sym
\T_{rt}(\Phi)$ is the multiplication, and $(\Delta')^{n}: \Sym
\T_{rt}(\Phi) \rightarrow \Sym \T_{rt}(\Phi)^{\o
  (n+1)}$ is the iterated application of the coassociative
$\Delta'$).
This extends to products of trees anti-multiplicatively.  So, it is
enough to check the bialgebra condition above.

\section{Quiver version and results}
We now proceed to the quiver versions of the preceding and formulate
our results.
\subsection{Necklace (pre-)Lie coalgebras}

\subsubsection{Original necklace Lie coalgebra (``oriented'')}
We recall the definition of the necklace Lie coalgebra from \cite{S},
which will correspond to the ``oriented'' case.
Let $Q$ be any quiver (with edge set also denoted by $Q$), and let
$\dq= Q \sqcup Q^*$ be the double quiver, $Q^* := \{e^*: e \in Q\}$,
where if $e$ is an arrow from $i$ to $j$ (denoted $e: i \rightarrow
j$),
then $e^*: j \rightarrow i$ is the reverse. The double quiver has the
same set $I$ of edges as $Q$.  Let $P$ be the path algebra on the
double quiver.  Precisely, one has $P = T_{\k^I} \langle \dq \rangle$,
where $\langle \dq \rangle$ is the $\k^I$-bimodule with basis $\dq$,
so that $i \langle \dq \rangle j$ is the space with basis those edges
$e: i \rightarrow j$.  For each edge $e: i \rightarrow j$ let $e_s := i,
e_t := j$ (the ``source'' and ``target,'' respectively).

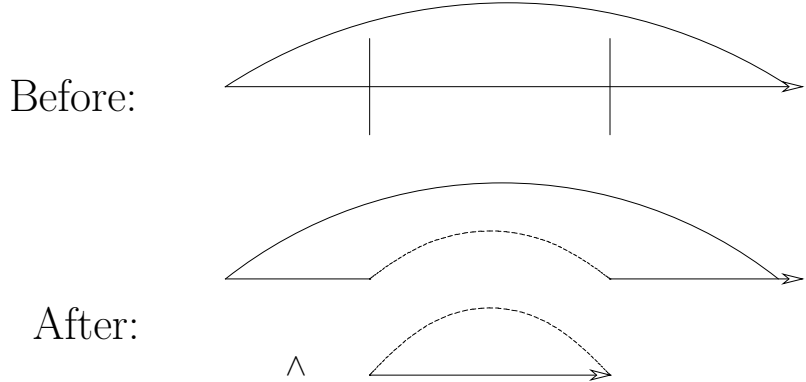
\begin{figure}[hbt]
\begin{center}
\setlength{\unitlength}{0.00083333in}
\begingroup\makeatletter\ifx\SetFigFont\undefined%
\gdef\SetFigFont#1#2#3#4#5{%
  \reset@font\fontsize{#1}{#2pt}%
  \fontfamily{#3}\fontseries{#4}\fontshape{#5}%
  \selectfont}%
\fi\endgroup%
{\renewcommand{\dashlinestretch}{30}
\begin{picture}(4977,2428)(0,-10)
\put(3127.500,-816.643){\arc{6439.723}{4.1332}{5.2916}}
\path(2265,2178)(2265,1578)
\path(3765,2178)(3765,1578)
\path(1365,1878)(4965,1878)
\whiten\path(4845.000,1848.000)(4965.000,1878.000)(4845.000,1908.000)(4881.000,1878.000)(4845.000,1848.000)
\put(15,1728){\makebox(0,0)[lb]{\smash{{\SetFigFont{17}{20.4}{\rmdefault}{\mddefault}{\updefault}Before:}}}}
\put(3090.000,-1501.688){\arc{5559.376}{4.0429}{5.3819}}
\path(2265,78)(3765,78)
\whiten\path(3645.000,48.000)(3765.000,78.000)(3645.000,108.000)(3681.000,78.000)(3645.000,48.000)
\path(1365,678)(2265,678)
\path(3765,678)(4965,678)
\whiten\path(4845.000,648.000)(4965.000,678.000)(4845.000,708.000)(4881.000,678.000)(4845.000,648.000)
\dashline{60.000}(2265,678)(2266,679)(2268,681)
	(2272,684)(2279,689)(2288,696)
	(2300,705)(2315,716)(2332,728)
	(2352,743)(2374,758)(2399,775)
	(2425,792)(2453,809)(2482,827)
	(2512,844)(2544,861)(2577,878)
	(2611,894)(2647,909)(2685,923)
	(2725,935)(2768,947)(2812,957)
	(2860,966)(2910,972)(2962,977)
	(3015,978)(3068,977)(3120,972)
	(3170,966)(3218,957)(3262,947)
	(3305,935)(3345,923)(3383,909)
	(3419,894)(3453,878)(3486,861)
	(3518,844)(3548,827)(3578,809)
	(3605,792)(3631,775)(3656,758)
	(3678,743)(3698,728)(3715,716)
	(3730,705)(3742,696)(3751,689)
	(3758,684)(3762,681)(3764,679)(3765,678)
\dashline{60.000}(2265,78)(2266,79)(2268,82)
	(2272,86)(2279,93)(2288,103)
	(2300,115)(2315,131)(2332,148)
	(2352,168)(2374,190)(2399,213)
	(2425,237)(2453,262)(2482,286)
	(2512,311)(2544,335)(2577,358)
	(2611,380)(2647,401)(2685,420)
	(2725,438)(2768,455)(2812,469)
	(2860,481)(2910,490)(2962,496)
	(3015,498)(3068,496)(3120,490)
	(3170,481)(3218,469)(3262,455)
	(3305,438)(3345,420)(3383,401)
	(3419,380)(3453,358)(3486,335)
	(3518,311)(3548,286)(3578,262)
	(3605,237)(3631,213)(3656,190)
	(3678,168)(3698,148)(3715,131)
	(3730,115)(3742,103)(3751,93)
	(3758,86)(3762,82)(3764,79)(3765,78)
\put(165,303){\makebox(0,0)[lb]{\smash{{\SetFigFont{17}{20.4}{\rmdefault}{\mddefault}{\updefault}After:}}}}
\put(1740,78){\makebox(0,0)[lb]{\smash{{\SetFigFont{12}{14.4}{\rmdefault}{\mddefault}{\updefault}$\wedge$}}}}
\end{picture}
}
\caption{The original Lie cobracket on $P/[P,P]$ from \cite{S}}
\label{lie-fig}
\end{center}
\end{figure}

Let $L := P/[P,P] = HH_0(P)$ be the $\k$-module with basis the cyclic
paths in the quiver $\dq$ (forgetting the initial edge).  We call such
cyclic paths ``necklaces,'' and the cobracket operation will involve
splitting necklaces into two necklaces (by making two cuts and gluing the
endpoints of the two resulting strands=paths in the quiver).
Then, one defines the cobracket $\delta = \delta_{or}: L \rightarrow L \wedge L$ (or=oriented)
as follows:
\begin{equation} \label{nlcd} \delta_{or}([a_1 \cdots a_n]) = \sum_{i
    < j}
  \omega(a_i, a_j) [(a_j)_t a_{j+1} \cdots a_{i-1}] \wedge [(a_i)_t
  a_{i+1} \cdots a_{j-1}],
\end{equation}
where $\omega(e, e^*) = -\omega(e^*, e) = 1$ for $e \in Q$, and $\omega(e, f) = 0$
if $e \neq f^*$ (we use the notation $(e^*)^* := e$).   A typical summand
is depicted in Figure \ref{lie-fig}.

\subsubsection{Necklace (pre-)Lie coalgebra of paths, ``rooted''}
We define ``rooted'' versions of the necklace Lie coalgebra.  By being
rooted, we will actually obtain a pre-Lie structure, as in the rooted tree case.

To add a basepoint to a necklace,  one should pick an initial edge. Equivalently, one can replace the necklace with a closed path. In this generality, one may actually speak of non-closed paths as well, which we do.

\begin{figure}[hbt]
\begin{center}
\setlength{\unitlength}{0.00083333in}
\begingroup\makeatletter\ifx\SetFigFont\undefined%
\gdef\SetFigFont#1#2#3#4#5{%
  \reset@font\fontsize{#1}{#2pt}%
  \fontfamily{#3}\fontseries{#4}\fontshape{#5}%
  \selectfont}%
\fi\endgroup%
{\renewcommand{\dashlinestretch}{30}
\begin{picture}(4977,1896)(0,-10)
\path(2265,1869)(2265,1269)
\path(3765,1869)(3765,1269)
\path(2265,69)(3765,69)
\whiten\path(3645.000,39.000)(3765.000,69.000)(3645.000,99.000)(3681.000,69.000)(3645.000,39.000)
\path(1365,669)(2265,669)
\path(3765,669)(4965,669)
\whiten\path(4845.000,639.000)(4965.000,669.000)(4845.000,699.000)(4881.000,669.000)(4845.000,639.000)
\path(1365,1569)(4965,1569)
\whiten\path(4845.000,1539.000)(4965.000,1569.000)(4845.000,1599.000)(4881.000,1569.000)(4845.000,1539.000)
\dashline{60.000}(2265,669)(2266,670)(2268,672)
	(2272,675)(2279,680)(2288,687)
	(2300,696)(2315,707)(2332,719)
	(2352,734)(2374,749)(2399,766)
	(2425,783)(2453,800)(2482,818)
	(2512,835)(2544,852)(2577,869)
	(2611,885)(2647,900)(2685,914)
	(2725,926)(2768,938)(2812,948)
	(2860,957)(2910,963)(2962,968)
	(3015,969)(3068,968)(3120,963)
	(3170,957)(3218,948)(3262,938)
	(3305,926)(3345,914)(3383,900)
	(3419,885)(3453,869)(3486,852)
	(3518,835)(3548,818)(3578,800)
	(3605,783)(3631,766)(3656,749)
	(3678,734)(3698,719)(3715,707)
	(3730,696)(3742,687)(3751,680)
	(3758,675)(3762,672)(3764,670)(3765,669)
\dashline{60.000}(2265,69)(2266,70)(2268,73)
	(2272,77)(2279,84)(2288,94)
	(2300,106)(2315,122)(2332,139)
	(2352,159)(2374,181)(2399,204)
	(2425,228)(2453,253)(2482,277)
	(2512,302)(2544,326)(2577,349)
	(2611,371)(2647,392)(2685,411)
	(2725,429)(2768,446)(2812,460)
	(2860,472)(2910,481)(2962,487)
	(3015,489)(3068,487)(3120,481)
	(3170,472)(3218,460)(3262,446)
	(3305,429)(3345,411)(3383,392)
	(3419,371)(3453,349)(3486,326)
	(3518,302)(3548,277)(3578,253)
	(3605,228)(3631,204)(3656,181)
	(3678,159)(3698,139)(3715,122)
	(3730,106)(3742,94)(3751,84)
	(3758,77)(3762,73)(3764,70)(3765,69)
\put(15,1419){\makebox(0,0)[lb]{\smash{{\SetFigFont{17}{20.4}{\rmdefault}{\mddefault}{\updefault}Before:}}}}
\put(165,294){\makebox(0,0)[lb]{\smash{{\SetFigFont{17}{20.4}{\rmdefault}{\mddefault}{\updefault}After:}}}}
\put(1740,69){\makebox(0,0)[lb]{\smash{{\SetFigFont{12}{14.4}{\rmdefault}{\mddefault}{\updefault}$\otimes$}}}}
\end{picture}
}
\caption{The Loday cobracket $P \rightarrow P/[P,P] \o P$ from \cite{S2}}
\label{loday-fig}
\end{center}
\end{figure}
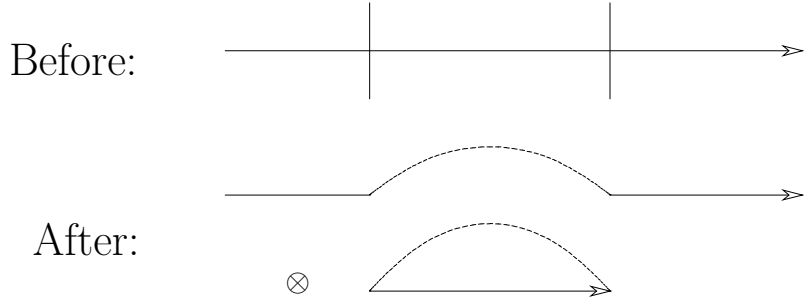

In \cite{S2}, a (Loday or Lie) cobracket is defined in this way by the
following idea: When one makes two cuts in a path and joins the cut
ends the same way we did before for necklaces, one obtains one path
and one necklace. This is depicted in Figure \ref{loday-fig}.
  
\begin{figure}
\begin{center}
\setlength{\unitlength}{0.00083333in}
\begingroup\makeatletter\ifx\SetFigFont\undefined%
\gdef\SetFigFont#1#2#3#4#5{%
  \reset@font\fontsize{#1}{#2pt}%
  \fontfamily{#3}\fontseries{#4}\fontshape{#5}%
  \selectfont}%
\fi\endgroup%
{\renewcommand{\dashlinestretch}{30}
\begin{picture}(4977,1896)(0,-10)
\path(2265,1869)(2265,1269)
\path(3765,1869)(3765,1269)
\path(2265,69)(3765,69)
\whiten\path(3645.000,39.000)(3765.000,69.000)(3645.000,99.000)(3681.000,69.000)(3645.000,39.000)
\path(1365,669)(2265,669)
\path(3765,669)(4965,669)
\whiten\path(4845.000,639.000)(4965.000,669.000)(4845.000,699.000)(4881.000,669.000)(4845.000,639.000)
\path(1365,1569)(4965,1569)
\whiten\path(4845.000,1539.000)(4965.000,1569.000)(4845.000,1599.000)(4881.000,1569.000)(4845.000,1539.000)
\dashline{60.000}(2265,669)(2266,670)(2268,672)
	(2272,675)(2279,680)(2288,687)
	(2300,696)(2315,707)(2332,719)
	(2352,734)(2374,749)(2399,766)
	(2425,783)(2453,800)(2482,818)
	(2512,835)(2544,852)(2577,869)
	(2611,885)(2647,900)(2685,914)
	(2725,926)(2768,938)(2812,948)
	(2860,957)(2910,963)(2962,968)
	(3015,969)(3068,968)(3120,963)
	(3170,957)(3218,948)(3262,938)
	(3305,926)(3345,914)(3383,900)
	(3419,885)(3453,869)(3486,852)
	(3518,835)(3548,818)(3578,800)
	(3605,783)(3631,766)(3656,749)
	(3678,734)(3698,719)(3715,707)
	(3730,696)(3742,687)(3751,680)
	(3758,675)(3762,672)(3764,670)(3765,669)
\put(15,1419){\makebox(0,0)[lb]{\smash{{\SetFigFont{17}{20.4}{\rmdefault}{\mddefault}{\updefault}Before:}}}}
\put(165,294){\makebox(0,0)[lb]{\smash{{\SetFigFont{17}{20.4}{\rmdefault}{\mddefault}{\updefault}After:}}}}
\put(1740,69){\makebox(0,0)[lb]{\smash{{\SetFigFont{12}{14.4}{\rmdefault}{\mddefault}{\updefault}$\otimes$}}}}
\end{picture}
}
\caption{A summand in the pre-Lie multiplication of ``Before''.}
\label{prelie-fig}
\end{center}
\end{figure}
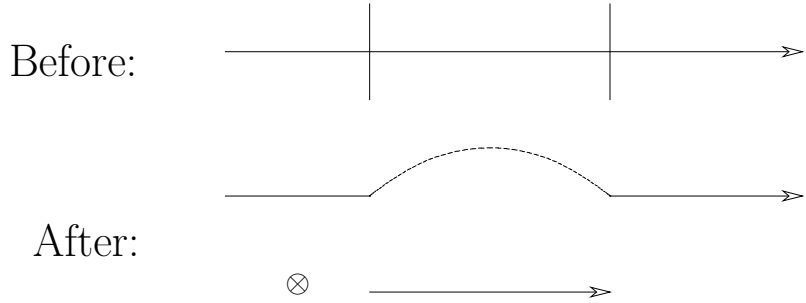

This will not give a pre-Lie coalgebra, however: to get one, one needs
(as in \cite{Tu2}) to split a path into two paths (resp. a rooted tree
into two rooted trees by cutting).  To do this, we make two cuts in
the path, but only glue once: the left strand to the right, obtaining
two strands (Figure \ref{prelie-fig}).
Precisely, we define the rooted pre-Lie comultiplication
$\delta_{p,rt}$ and Lie cobracket $\delta_{rt}$ by the formulas
\begin{gather} \label{prte}
\delta_{p,rt}(a_1 \cdots a_n) = \sum_{i < j} -\omega(a_i, a_j) (a_i)_t a_{i+1} \cdots a_{j-1} \o
(a_1)_s a_1 \cdots a_{i-1} a_{j+1} \cdots a_n, \\
\delta_{rt} = \delta_{p,rt} - \tau_{(12)} \delta_{p,rt}.
\end{gather}
\begin{prop} The maps $\delta_{or}, \delta_{rt}$ are Lie cobrackets, and
  $\delta_{p,rt}$ is a pre-Lie comultiplication.
\end{prop}
\begin{proof}
  It suffices to check that $\delta_{p, rt}$ is a pre-Lie
  comultiplication ($\delta_{or}$ is a Lie cobracket by \cite{S}).
  This follows along similar lines to the proof that $\delta_{or}$ is
  a Lie cobracket in \cite{S}, \S 2.2.
\end{proof}
Note that, while in the rooted case, one may consider paths that are
not closed, the first component of the image of $\delta_{p,rt}$ lies
in the span of closed paths.

\subsection{Hopf algebra of paths}

We now define a Hopf algebra which completes the analogy ``pre-Lie
coalgebra of rooted trees :: Renormalization Hopf algebra == Necklace
pre-Lie coalgebra of paths (rooted) :: ??''  Note that this does not have an oriented version (since there is no pre-Lie coalgebra in the oriented case, cf.~Proposition \ref{preleqhop}).
\begin{defn}
  Given a path $a_1 \cdots a_n \in P$, a \textit{cut} $H$ is a
  choice of pairs
\begin{equation}
H=\{(i_1,j_1), \ldots, (i_m, j_m)\} \subset \{1,\ldots,n\}^{2},
\end{equation} such that
\begin{enumerate}
\item $i_1, j_1, \ldots, i_m, j_m$ are distinct,
\item for all $\ell$, $i_\ell < j_\ell$,
\item the pairs do not cross: that is, there do not exist
  $\ell, \ell'$ such that $i_\ell < i_{\ell'} < j_\ell < j_{\ell'}$, and
\item for all $\ell$, there exists $e_\ell \in Q$ such that
  $\{a_{i_\ell},a_{j_\ell}\} = \{e_\ell,e_\ell^*\}$.
\end{enumerate}
\end{defn}
\begin{defn}
A cut is called \textbf{simple} if there do not exist $\ell,\ell'$ with $i_\ell < i_{\ell'} < j_{\ell'} < j_{\ell}$: that is, as in Figure \ref{dualtree-fig}, no added (semicircular) edge contains another such.
\end{defn}
\begin{defn} For any cut $H$ of a path $X := a_1 \cdots a_n$, let
  $\{X_{H,c}\}_{c \in H} \sqcup \{X_{H,0}\}$ be the collection of
  paths obtained by applying Figure \ref{prelie-fig} repeatedly (cut
  each pair of edges $a_{i_\ell}, a_{j_\ell}$ and glue one pair of
  endpoints each, as in \eqref{prte}). By definition, $X_{H,0}$ is the
  unique path which shares the endpoints of the original path
  (beginning at $(a_1)_s$ and ending at $(a_n)_t$), and $X_{H,c}$ for
  $c = (i_\ell,j_\ell)$ is the unique path which begins at the target
  $(a_{i_\ell})_t$ of $a_{i_\ell}$, and ends at the source
  $(a_{j_\ell})_s$ of $a_{j_\ell}$.
\end{defn}
\begin{defn}
  For any cut $H = \{(i_1, j_1), \ldots, (i_m, j_m)\}$ of a path
  $a_1 \cdots a_n$, let $\varepsilon_H = \pm 1$ be defined by
\begin{equation}
\varepsilon_H := \prod_{\ell=1}^m -\omega(a_{i_\ell}, a_{j_\ell}).
\end{equation}
\end{defn}
\begin{defn} Define the coproduct $\Delta: \Sym P \rightarrow (\Sym P)^{\otimes 2}$ on
a path $X = a_1 \cdots a_n$ by
\begin{gather} \label{lhdfn}
l_H(X) := X_{H,c_1} \& \cdots \& X_{H,c_{|H|}}, \quad r_H(X) := X_{H,0}, \\
\Delta(X) := X \o 1 + \sum_{\text{simple cuts } H} \varepsilon_H l_H(X) \o r_H(X), \label{deldfn}
\end{gather}
where $H = \{c_1, \ldots, c_{|H|}\}$.
\end{defn}
\begin{prop} \label{sympishopf} The map $\Delta$ endows $\Sym P$ with the structure of a commutative
Hopf algebra with antipode given by \eqref{afla} (where $X$ is a path).
\end{prop}
This proposition will be proved in Section \ref{sympishopfs}.  Note
that the result also follows from Theorem \ref{homt1}, since the map
$\eta$ gives a monomorphism of Hopf algebras (i.e., without knowing
$\Sym P$ is Hopf, the theorem shows that $\eta$ is injective and
carries the proposed multiplication, comultiplication, unit, and
counit to those for Connes and Kreimer's Hopf algebra).

\subsection{The monomorphisms to renormalization algebras}
\subsubsection{The tree-structures}
Following in analogy with \cite{Tu2}, we define a pre-Lie coalgebra map
$(P,\delta_{p,rt}) \rightarrow (\T_{rt}(\Phi_{rt}), \rho)$ and
a Lie coalgebra map 
$(L, \delta_{or}) \rightarrow (\T_{or}(\Phi_{or}), \rho^{ss}_O)$ for
certain rooted (resp. oriented) tree-structures $\Phi_{rt},
\Phi_{or}$.
The former induces a Hopf algebra map
$(\Sym(P),\Delta) \rightarrow (\Sym(\T_{rt}), \Delta)$, as we explain.
As a result of the pre-Lie coalgebra map, one also obtains a Lie coalgebra
map on the associated Lie coalgebra $(P, \delta_{rt})$ 
(by skew-symmetrizing the
pre-Lie comultiplication).

\begin{defn}
A \textbf{ribbon graph structure} on a tree is a choice, for each vertex of the
tree, of a fixed cyclic ordering of the edges incident with that vertex.
\end{defn}
\begin{defn}
Let $\Phi_{or}$ be the oriented tree-structure which assigns to each oriented
tree a choice of ribbon graph structure on the tree, and to each vertex of the
tree a cyclic path in $\dq$ (which is a basis element of $L$). 
\end{defn}
\begin{defn}\cite{Tu2}
  A \textbf{corner}\footnote{This is called a \textbf{corner} to agree
    with \cite{Tu2}, Remark 2 of \S 5.2, where it is defined as a
    choice of two consecutive edges in the cyclic ordering (these are
    the last and first edges in our linear ordering).}
  of a vertex of a ribbon graph is a choice of initial edge at the vertex (giving a
  linear ordering of the incident edges).
\end{defn}
\begin{defn}
  Let $\Phi_{rt}$ be the rooted tree-structure which assigns to each
  rooted tree a choice of ribbon graph structure with a corner at the
  root, together with an orientation of all edges, and a
  labeling of vertices by paths in $\dq$ (basis elements of $P$).  
\end{defn}
\begin{rem}
Instead of assigning (cyclic) paths to each vertex, an alternative would be
to assign a single element of $P^{\o V}$ or $L^{\o V}$ to the tree, where
$\o V$ means taking $\o |V|$ with components labeled by $V$. Then, $\Phi_*$ would obtain a $\k$-module structure, and we could work with the quotient
$\overline{\T(\Phi_*)}$ of $\T(\Phi_*)$ by the relation $(T, s + s') = (T,s) + (T,s')$.
\end{rem}
\begin{rem}
  Note that, at all vertices other than the root of a rooted tree with
  a ribbon-graph structure, the cyclic ordering actually has a
  canonical lifting to a linear ordering (i.e., a corner), by choosing
  as initial the edge that lies between the given vertex and the root.
  So with the corner at the root, one obtains rooted trees with linear
  orderings at all vertices (rather than merely cyclic orderings).
\end{rem}
\subsubsection{Chord diagrams and dual trees}
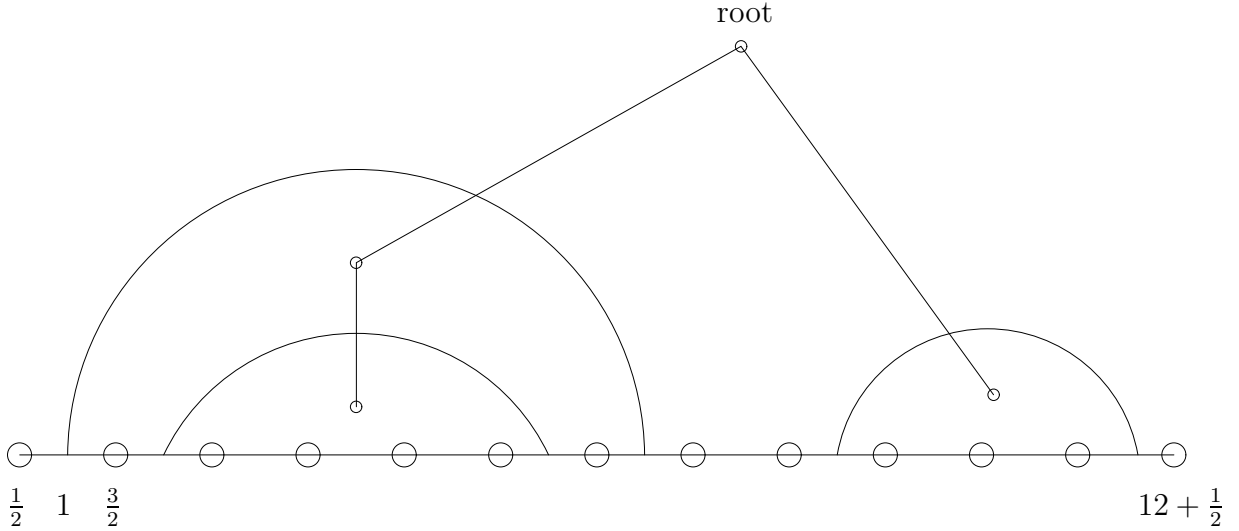
\begin{figure}
\begin{center}
\setlength{\unitlength}{0.00083333in}
\begingroup\makeatletter\ifx\SetFigFont\undefined%
\gdef\SetFigFont#1#2#3#4#5{%
  \reset@font\fontsize{#1}{#2pt}%
  \fontfamily{#3}\fontseries{#4}\fontshape{#5}%
  \selectfont}%
\fi\endgroup%
{\renewcommand{\dashlinestretch}{30}
\begin{picture}(7373,3303)(0,-10)
\put(2190.000,435.065){\arc{3600.179}{3.1516}{6.2732}}
\put(2190.000,-117.000){\arc{2656.991}{3.5850}{5.8397}}
\put(6127.500,288.000){\arc{1903.819}{3.3158}{6.1090}}
\put(690,453){\ellipse{150}{150}}
\put(1290,453){\ellipse{150}{150}}
\put(1890,453){\ellipse{150}{150}}
\put(2490,453){\ellipse{150}{150}}
\put(3090,453){\ellipse{150}{150}}
\put(3690,453){\ellipse{150}{150}}
\put(4290,453){\ellipse{150}{150}}
\put(4890,453){\ellipse{150}{150}}
\put(5490,453){\ellipse{150}{150}}
\put(6690,453){\ellipse{150}{150}}
\put(6090,453){\ellipse{150}{150}}
\put(7290,453){\ellipse{150}{150}}
\put(90,453){\ellipse{150}{150}}
\put(4590,3003){\ellipse{72}{72}}
\put(6165,828){\ellipse{72}{72}}
\put(2190,753){\ellipse{72}{72}}
\put(2190,1653){\ellipse{72}{72}}
\path(90,453)(7290,453)
\path(2190,753)(2190,1653)(4590,3003)(6165,828)
\put(4440,3153){\makebox(0,0)[lb]{\smash{{\SetFigFont{12}{14.4}{\rmdefault}{\mddefault}{\updefault}root}}}}
\put(15,78){\makebox(0,0)[lb]{\smash{{\SetFigFont{12}{14.4}{\rmdefault}{\mddefault}{\updefault}$\frac{1}{2}$}}}}
\put(7065,78){\makebox(0,0)[lb]{\smash{{\SetFigFont{12}{14.4}{\rmdefault}{\mddefault}{\updefault}$12+\frac{1}{2}$}}}}
\put(315,78){\makebox(0,0)[lb]{\smash{{\SetFigFont{12}{14.4}{\rmdefault}{\mddefault}{\updefault}$1$}}}}
\put(615,78){\makebox(0,0)[lb]{\smash{{\SetFigFont{12}{14.4}{\rmdefault}{\mddefault}{\updefault}$\frac{3}{2}$}}}}
\end{picture}
}
\end{center}
\caption{A typical chord diagram (for a cut) and its dual tree.}
\label{dualtree-fig}
\end{figure}

In order to define the homomorphisms of pre-Lie coalgebras and Hopf
algebras, we first need to construct from a simple cut of a path (or
later, a cyclic path), the dual tree to the chord diagram associated
to this cut:
\begin{defn}
  For any path $X = a_1 \cdots a_n$ of length $n$, associate to this a
  line segment $L_X \subset \R$ with edges
  $e_i = [i-\frac{1}{2}, i+\frac{1}{2}]$ for all $1 \leq i \leq n$,
  and vertices $\{\frac{1}{2}, \frac{3}{2}, \ldots, n+\frac{1}{2}\}$.
  The \textit{root} is defined to be $rt := \frac{1}{2}$. (See Figure
  \ref{dualtree-fig}).
\end{defn}
We also think of the vertex $n+ \frac{1}{2}$ as the root, essentially
considering it to be the same vertex as $\frac{1}{2}$.  That is, a
circle with basepoint is essentially equivalent to a line by cutting
at the basepoint: then, the endpoints of the line are both the
basepoint. We chose $\frac{1}{2}$ rather than $n+\frac{1}{2}$ for the
root only for definiteness: the choice makes no difference.
\begin{defn}
  For any cut $H = \{(i_1, j_1), \ldots, (i_m,j_m)\}$ of a path
  $X = a_1 \cdots a_n \in P$, consider the associated \textbf{chord
    diagram}
  $C_{X,H}$, obtained from $L_X$ by adding interior vertices
  $i_\ell, j_\ell$ to the edges $e_{i_\ell}, e_{j_\ell}$, and a new
  edge with endpoints $i_\ell, j_\ell$ for each $\ell$, as in Figure
  \ref{dualtree-fig}.  The edges are chosen so as to not intersect
  (giving a planar graph).
\end{defn}
\begin{defn}
  For any cut $H$ as above, let the \textbf{dual rooted tree} $T_H$
be obtained by dualizing the chord
diagram: place one vertex inside each face of the chord diagram, and one edge crossing each
edge of the chord diagram, connecting the vertices associated to the two faces.
The root corresponds to the unbounded face (which is included as a face).
\end{defn}
\begin{defn} For any cut $H$ of a path as above with dual rooted tree
  $T_H$, we define an element $s_H$ of $\Phi_{rt}(T_H)$ as follows:
  First, orient the edges of the chord diagram, by assigning the edge
  with endpoints $i_\ell, j_\ell$ the orientation
  $i_\ell \rightarrow j_\ell$ if $a_{i_\ell} \in Q$ and
  $j_\ell \rightarrow i_\ell$ otherwise. Then, the orientation of the
  edge $e$ of $T_H$ which crosses an edge $f$ of the chord diagram is
  such that $e \wedge f$ is the positive orientation on $\R^2$.  Next,
  the unbounded face is still considered a face, and its vertex is
  declared the root.  This is naturally a ribbon graph. The linear
  ordering of the edges at the root (choice of corner) is given by
  the usual linear ordering of the endpoints of the edges of the
  unbounded face of the chord diagram in the interval
  $[\frac{1}{2},n+\frac{1}{2}]$.  Finally, the labeling of the
  vertices is given as follows: to each face $f$ of the chord diagram,
  $J = \partial f \cap [\frac{1}{2},n+\frac{1}{2}]$ is a union of
  closed intervals; let $J^{\circ}$ be the interior and
  $I^{\circ} \cap \{1,2,\ldots,n\}$ the set of integers incident to
  the face which are not incident to any of the edges of the chord
  diagram (not including $[\frac{1}{2},n+\frac{1}{2}]$). Let $i_J$ be
  the vertex of the quiver which is the left endpoint of $J$: that is,
  $i_J = (a_{\min(J \cap \{1,\ldots,n\})})_t$. Then,
  $i_{J} \prod_{j \in J^{\circ} \cap \{1,2,\ldots,n\}} a_j$ is the path
  associated to the vertex attached to $f$ (it is the path which
  remains in that face after performing cuts as in Figure
  \ref{dualtree-fig}).
\end{defn}
The dual tree is depicted (without the tree-structure) in Figure \ref{dualtree-fig}.
We proceed to define the dual oriented tree:
\begin{defn} For any cut $H$ as above, the \textbf{dual oriented tree}  
is the dual tree $T_H$, forgetting the root, together with the orientation
of edges given by the element of $\Phi_{rt}(T_H)$ above. Call it $T_{H,or}$.
\end{defn}
\begin{defn} For any closed path $a_1 \cdots a_m$ with dual oriented tree $T_{H,or}$, 
define an element $s_{H,or}$ of $\Phi_{or}(T_{H,or})$ from $s_H$
by taking the image of the labels of vertices in cyclic paths ($L=P/[P,P]$),
and forgetting the corner structure.
\end{defn}
One may easily verify the
\begin{lemma}
  For any closed path $a_1 \cdots a_m$, the space of cuts of
  $a_1 \cdots a_m$ is naturally isomorphic to that of
  $a_i a_{i+1} \cdots a_{i-1}$ for all $i$, in a way that yields a
  natural isomorphism of dual oriented trees which carries the elements $\Phi_{or}(T_{H,or})$
  to each other. One may therefore define a

  cut of the cyclic path $[a_1 \cdots a_m]$ and its associated
  dual oriented tree $T_{H,or}$, with associated element $s_{H,or} \in \Phi_{or}(T_{H,or})$.
\end{lemma}
\subsubsection{The homomorphisms} \label{homs} 
Finally, we may define the homomorphisms and state the first theorem.
\begin{defn}
For any path $a_1 \cdots a_n$, define the element $\eta_{rt}(a_1 \cdots a_n) \in \T_{rt}(\Phi_{rt})$
as follows: 
\begin{equation}
\eta_{rt}(a_1 \cdots a_n) = \sum_{\text{cuts } H} \varepsilon_H (T_{H}, s_H).
\end{equation}
\end{defn}
\begin{defn}
  For any cyclic path $[a_1 \cdots a_n]$, define the element
  $\eta_{or}([a_1 \cdots a_n]) \in \T_{or}(\Phi_{or})$ as follows:
\begin{equation}
\eta_{or}([a_1 \cdots a_n]) = \sum_{\text{cuts } H} (T_{H,or}, s_{H,or}).
\end{equation}
\end{defn}
\begin{thm} \label{homt1} The maps $\eta_{or}$ and $\eta_{rt}$ extend
  linearly to an injective pre-Lie coalgebra homomorphism
  $(P, \delta_{p, rt}) \rightarrow \T_{rt}(\Phi_{rt})$
  and an injective Lie coalgebra homomorphism
  $(L,\delta_{or}) \rightarrow \T_{or}(\Phi_{or})$, respectively.
  Furthermore, $\eta_{rt}$ extends multiplicatively (and linearly) to
  a Hopf algebra monomorphism
  $\Sym P \rightarrow \Sym \T_{rt}(\Phi_{rt})$.
\end{thm}
This theorem will be proved in Section \ref{homt12ps}.

M. Livernet pointed out to us that, by \cite{CL}, the pre-Lie coalgebra
of decorated trees $\T_{rt}(\Phi)$ is a cofree pre-Lie coalgebra, for any
$\Phi$. This explains why morphisms such as the above must always
exist (although the one we construct is particularly natural).

\subsection{Factorization of $\eta$ through chord algebras}
It turns out that one can understand the $\eta$ homomorphisms (and
their injectivity) through a factorization as follows:
\begin{gather} \label{chorf}
(L,\delta_{or}) \mathop{\into}^{S_{or}} 
Ch_{or} \tra^{D_{or}} \T_{or}(\Phi_{or}), \\
\label{chrtf} 
(P,\delta_{p,rt}) \mathop{\into}^{S_{rt}} Ch_{rt} 
\tra^{D_{rt}} \T_{rt}(\Phi_{rt}),
\end{gather}
where $Ch_{or}$ and $Ch_{rt}$, called \textit{chord algebras}, are
spanned by chord diagrams on necklaces or paths, and the first maps in
\eqref{chorf},\eqref{chrtf} take a necklace (resp., path) to the sum
of all possible chord diagrams on that necklace or path.  We will
equip $Ch_{rt}$ and $Ch_{or}$ with the appropriate pre-Lie and Lie coalgebra
structures, and $\Sym Ch_{rt}$ with the appropriate Hopf algebra
structure, so that one obtains the following sequence of Hopf algebra
homomorphisms:
\begin{equation} \label{symchrtf}
\Sym P \mathop{\into}^{S_{rt}} \Sym Ch_{rt} \tra^{D_{rt}} \Sym \T_{rt}(\Phi_{rt}).
\end{equation}
We also briefly explain how this construction can also be done in the
context of \cite{Tu2}, yielding the space of geometric chord diagrams.

\begin{defn} A necklace chord diagram is a necklace (=cyclic monomial) $[a_1 \cdots a_n] \in L$, for $a_i \in \dq$, together with a cut $H$ of $[a_1 \cdots a_n]$. Denote the necklace chord diagram by $([a_1 \cdots a_n],H)$.
\end{defn}
\begin{defn} A path chord diagram is a path $a_1 \cdots a_n \in P$ for $a_i \in \dq$, together with a cut $H$ of $a_1 \cdots a_n$. The path chord diagram is
denoted by $(a_1 \cdots a_n, H)$.
\end{defn}
\begin{defn} The chord algebra $Ch_{rt}$ is defined(as a linear space)
  to be the free $\k$-module with basis the path chord diagrams.
  Similarly, $Ch_{or}$ is defined to have basis the necklace chord
  diagrams.
\end{defn}
\begin{defn} \label{chord}
Define $\delta_{p,rt}: Ch_{rt} \rightarrow Ch_{rt}^{\o 2}$ 
as follows. Let $X = a_1 \cdots a_n$ be a path, for
$a_\ell \in \dq$, and let $H = \{(i_1, j_1), \ldots, (i_m, j_m)\}$ be a
cut. For any $1 \leq \ell \leq m$, let $c_\ell := (i_\ell, j_\ell)$,
and let $H^1_{c_\ell}, H^2_{c_\ell} \subset H$
be the two subcuts obtained by removing $c_\ell=(i_\ell, j_\ell)$ from $H$:
$H^1_{c_\ell}$ is the collection of $(i_{\ell'}, j_{\ell'})$ on the
inside of $c_\ell$ (in particular $\ell' \neq \ell$),
and $H^2_{c_\ell}$ is the collection on the outside of
$c_\ell$.  Explicitly, $(i_{\ell'}, j_{\ell'}) \in H^1_{c_\ell}$
iff $i_{\ell'} > i_\ell$ (equivalently, $j_{\ell'} < j_\ell$). 
Finally, we then let $X_{c_\ell,1}, X_{c_\ell,2}$ be 
two chord diagrams thus
obtained: $X_{c_\ell,1} := ((a_{i_\ell})_t a_{i_\ell+1} \cdots a_{j_{\ell}-1}, H_{c_\ell}^1)$
and $X_{c_\ell,2} := (a_1 \cdots a_{i_\ell-1} (a_{j_\ell})_t a_{j_\ell+1} \cdots a_n, H_{c_\ell}^2)$.
Then, we
define 
\begin{equation}
\delta_{p,rt} (X, H) 
= \sum_{\ell=1}^{m} -\omega(a_{i_\ell}, a_{j_\ell}) X_{c_\ell,1} \o X_{c_\ell,2}.
\end{equation} 
\end{defn}
\begin{defn} \label{chrtd}
Define $\delta_{or}: Ch_{or} \rightarrow Ch_{or}^{\o 2}$ as follows.
Let $X = [a_1 \cdots a_n]$ be a necklace, for
$a_\ell \in \dq$, and let $H$ be a cut of $X$, which corresponds to 
the cut $\{(i_1, j_1), \ldots, (i_m, j_m)\}$ of $a_1 \cdots a_n$. 
For any $1 \leq \ell \leq m$, 
let $X_{c_\ell,1}, X_{c_\ell,2}$
be the two chord diagrams obtained by removing $(i_\ell, j_\ell)$ from $H$,
defined as in Definition \ref{chord}, except adding braces $[ \, ]$ around
the obtained paths.
Then, define
\begin{equation}
\delta_{or} (X, H) 
= \sum_{\ell=1}^{m} -\omega(a_{i_\ell}, a_{j_\ell}) (X_{c_\ell,1} \o X_{c_\ell, 2} - X_{c_\ell, 2} \o X_{c_\ell,1}).
\end{equation}
\end{defn}
\begin{defn}
Define the coproduct $\Delta$ on $\Sym Ch_{rt}$ as follows: For any
chord diagram $X := (a_1 \cdots a_n, H)$, with $H = \{(i_1,j_1),\ldots,(i_m,j_m)\}$, and any simple cut $H'$ with
$H' \subset H$, 
let $\{X_{H',c}\}_{c \in H'} \cup \{X_{H,0}\}$
be the collection of chord diagrams obtained by cutting out the
chords in $H'$: each time we cut out a chord from $H'$, we divide
a cut into two separate cuts, as in Definition \ref{chord}, and divide
the corresponding path into two paths. Then, $X_{H,0}$ is the obtained 
chord diagram which contains the basepoint ($(a_1)_s$ and $(a_n)_t$), and
$\{X_{H',c}\}$ is the other chord diagram which is cut from the chord
$c \in H'$.  Then, we define
\begin{gather}
l_{H'}(X) := \prod_{c \in H'} X_{H',c}, \quad r_{H'}(X) = X_{H',0}, \quad
\epsilon_{H'} = \prod_{c = (i_\ell, j_\ell) \in H} -\omega(a_{i_\ell},a_{j_\ell}),\\
\Delta((a_1 \cdots a_n, H)) = X \o 1 + \sum_{\text{simple cuts $H' \subset H$}} \epsilon_{H'} l_{H'}(X) \o r_{H'}(X). \label{chcod}
\end{gather}
\end{defn}
Finally, we have the following theorem, which is a strengthening of
Theorem \ref{homt1}.
\begin{thm} \label{homt2} The algebra $(Ch_{rt}, \delta_{p,rt})$ 
  is a pre-Lie coalgebra
  and $(Ch_{or},\delta_{or})$ is a Lie coalgebra,
  and 
  $\Sym Ch_{rt}$ is a Hopf algebra given by \eqref{chcod},
  \eqref{afla} (with counit given by $\epsilon(X)=0$ for any chord
  diagram $X$).  Then, the diagrams \eqref{chorf},\eqref{chrtf}, and
  \eqref{symchrtf} are homomorphisms, where $S_{*}$ takes a path or
  necklace to the sum over all chord diagrams over that path or
  necklace, and $D_*$ takes a chord diagram with cut $H$ to $\varepsilon_H$ times its dual tree,
  assigning data as in Section \ref{homs}.
\end{thm}
This theorem will be proved in Section \ref{homt12ps}.
\subsubsection{Geometric chord diagrams}
We briefly indicate the geometric counterpart of Theorem \ref{homt2}
(since algebras of chord diagrams were not discussed in \cite{Tu2}).
We consider geometric chord diagrams with a single loop: this means
(\cite{AMR}) a smooth map of a loop with chords into a surface, such
that the map is constant on the chords.  Then, we let $Ch_{rt}$ be the
algebra which, as a free $\k$-module, has basis the isotopy classes of
geometric chord diagrams with a single loop with basepoint (the
isotopies must be through such geometric chord diagrams), and $Ch_{or}$
is, as a $\k$-module, the space of isotopy classes of geometric chord
diagrams with a single loop without basepoint. One may then form the
sequences of homomorphisms \eqref{chorf}, \eqref{chrtf}, and
\eqref{symchrtf}: the map $S_*$ takes an isotopy class of loops to the
sum of all isotopy classes of chord diagrams whose underlying isotopy
class of loops is the original class, and the map $D_*$ takes a
geometric chord diagram to the dual tree, which then has all the
structure required of $\Phi_{or}, \Phi_{rt}$.

\subsection{Relationship between the Hopf algebras and pre-Lie algebras}
We note that the Hopf algebras considered here all have the following
special form: as an algebra, they are $\Sym V$ for some vector space $V$,
and the comultiplication $\Delta$ and counit $\epsilon$ have the form
\begin{equation} \label{hsf}
\Delta(v) = v \o 1 + 1 \o v + \Delta'(v), \quad \Delta'(v) \in (\Syns^{\geq 1}\  V) \o V, \quad \epsilon(v) = 0, \quad \forall v \in V.
\end{equation}
We claim that such Hopf algebras are in one-to-one correspondence with
pre-Lie comultiplications on the vector space $V$, as follows (this
was essentially observed in \cite{GuOu} in the dual setting, but not
quite formulated the same way):
\begin{prop} \label{preleqhop} Let $V$ be any $\Z_{+}$-graded vector space.
\begin{enumerate}
\item[(i)] For any Hopf algebra on $\Sym V$ satisfying \eqref{hsf}, the
map $\rho: V \rightarrow V \o V$ given by the composition of $\Delta'$ with
the projection to $V \o V$, is a pre-Lie comultiplication.
\item[(ii)] Conversely, given any pre-Lie comultiplication $\rho: V \rightarrow V \o V$ which preserves the total grading (induced by the grading on $V$), there exists a unique graded comultiplication on $\Sym V$ of the form \eqref{hsf}, which yields $\rho$ as in part (i).
\item[(iii)] Moreover, we describe an inductive procedure for computing
the $\Delta$ guaranteed in part (ii).
\end{enumerate}
\end{prop}
\begin{proof} In all parts, let subscripts $_n$  denote the standard (not total) grading on $\Sym V$, so that $(\Sym V)_n = \text{Sym}^n V$, 
extended to $(\Sym V)^{\o 2}$ and $(\Sym V)^{\o 3}$ by
\begin{equation}
(\Sym V)^{\o 2}_n = \bigoplus_{i+j=n} \text{Sym}^i V \o \text{Sym}^j V, \quad
(\Sym V)^{\o 3}_n = \bigoplus_{i+j+k=n} \text{Sym}^i V \o \text{Sym}^j V \o \text{Sym}^k V.
\end{equation}

  (i) Let us formally write
\begin{equation} \label{dexp}
\Delta = \Delta_0 + \Delta_1 + \Delta_2 + \ldots,
\end{equation} where
$\Delta_n : V \rightarrow \text{Sym}^{n} V \o V$ for $n \geq 1$ and
$\Delta_0(v) = 1 \o v + v \o 1$ for any $v \in V$ (and $\Delta_n(1) = \delta_{n,0} \cdot (1 \o 1)$);  for any
$v \in V$, only finitely many of the $\Delta_n(v)$ can be nonzero.  Also, let 
$\Delta_0'(X) := \Delta_0(X) - (1 \o X + X \o 1)$  (thus, $\Delta_0'(v_1 v_2) = v_1 \o v_2 + v_2 \o v_1$ for any $v_1, v_2 \in V$).  Then,
modulo $\bigoplus_{m \geq 4} (\Sym V)^{\o 3}_{m}$, one has
\begin{equation}
0 =  (\Delta \o 1) \Delta(v) - (1 \o \Delta) \Delta(v) = \bigl((\Delta_1 \o 1) \Delta_1(v) - (1 \o \Delta_1) \Delta_1(v)\bigr) + (\Delta_0' \o 1) \Delta_2(v).
\end{equation}
This equation says that, setting $\rho := \Delta_1$, $\rho$ is pre-Lie (since
$(\Delta_0' \o 1) \Delta_2(v)$ is symmetric in the first and second components).

(ii,iii) Suppose we are given $\Delta_1, \ldots, \Delta_n$ for
$n \geq 1$ such that
$\Delta_{\leq n} := \Delta_0 + \Delta_1 + \ldots + \Delta_n$, extended
multiplicatively to $\Sym V$, is coassociative modulo
$\bigoplus_{m \geq n+2} (\Sym V)^{\o 3}_m$.  We would like to find 
$\Delta_{n+1}$ such that $\Delta_{\leq n} + \Delta_{n+1}$ is coassociative modulo $\bigoplus_{m \geq n+3}(\Sym V)^{\o 3}_m$. This is equivalent to
\begin{equation}
  (\Delta_0' \o 1) \Delta_{n+1}(v) = \sum_{i+j=n+1, i,j > 0} 
(1 \o \Delta_i - \Delta_i \o 1) \Delta_j (v).
\end{equation}
In particular, $\Delta_{n+1}$ exists iff the first two components live in
$\Delta_0'(\Sym V)$, and in this case $\Delta_{n+1}$ is
unique, and may be computed algorithmically as indicated.  Existence
follows from the dual version of the construction of \cite{GuOu}. 
\end{proof}

We note that we used the grading above only to guarantee finiteness of
the sum $\Delta_0 + \Delta_1 + \cdots$ on any vector $v \in V$:
without assuming gradedness, the above proposition still holds if we
work in the completed tensor product $\Sym V \o \Sym V$ with respect
to the grading $(\Sym V)^{\o 2}_{\bullet}$.  Alternatively, one could
assume that iterated applications of $\rho$ on any $v \in V$
eventually yield zero.

As a result of the proposition, proving the main results of this paper
(or \cite{Tu2}) on the pre-Lie level is in fact equivalent to proving
them on the Hopf algebra level, as one can translate between the two using
the above proposition.  This explains why one must arrive at
\eqref{deldfn} (resp., the formulas from \cite{Tu2}, \S 8.3) for
coproduct given the choice of pre-Lie comultiplication.

We see that the fact that the pre-Lie structure fails to exist in the
oriented case is the same as the fact that the renormalization Hopf
algebra does not exist without using rooted trees.

\subsubsection{Noncommutative version}
In \cite{Tu2}, \S 8.5, a ``noncommutative'' version of the Hopf
algebras was defined, using the tensor algebras over $V$ instead of
the symmetric algebras.  In this version, $\Delta_0$ is the usual
``shuffle'' coproduct on $T(V)$ generated multiplicatively by
$\Delta_0(v) = 1 \o v + v \o 1$ for $v \in V$.  In this case,
Proposition \ref{preleqhop} is no longer true (for instance,
$\Delta'_0(vw-wv) = 0$ for any $v, w \in V$, so that one would have a
choice of $\Delta_2$).

However, in the case of paths, as in Turaev's case of loops, one has a
canonical choice of $\Delta$.  Namely, in $T(P)$, we can define the coproduct
by \eqref{lhdfn},\eqref{deldfn} except replacing $l_H$ \eqref{lhdfn}
by an ordered tensor product, choosing the left-to-right order of
components induced by the original ordering on the path.  As in
\cite{Tu2}, it is easy to check that this gives a Hopf algebra, and
that a suitable version of the homomorphism $\eta$ maps this Hopf
algebra to Foissy's noncommutative algebra of labeled rooted trees \cite{F}
(the rooted trees must have labeled edges, or equivalently, labeled
vertices, in order to capture the left-to-right order that we obtain
in paths).  So more ``noncommutative'' or ``ordering'' information is
included in paths or loops than in rooted trees without labels.

  On the other hand, one can obtain a generalization of Proposition
  \ref{preleqhop} if one imposes the additional condition that
  $\Delta_n(V) \subset \text{Sym}^n V \o V \subset T^n(V) \o V$.  In this
  case, there is not really anything new to check since
  coassociativity is proved in $\Sym(V) \subset T(V)$ and extends to
  all of $T(V)$ using the bialgebra condition (since we define
  $\Delta(fg) := \Delta(f)\Delta(g)$).  This might be
  the dual version of the noncommutative structure hinted at in Remark 2.14 of \cite{GuOu}. However, this is not the way
  to construct the noncommutative Hopf algebras described above, since it uses
  extra structure (the ordering in paths, or in the case of trees, the
  labeling on edges and/or vertices), and does not map $V$ to $(\Sym V \o V)
\oplus (1 \o V)$.
\section{Postponed proofs}
\subsection{Proof of Proposition \ref{sympishopf}} \label{sympishopfs}
  We only need to check coassociativity. Then, the bialgebra property
  essentially follows from the definition.  Coassociativity is easily 
verified by an explicit formula as follows:
\begin{defn}
For any path $X = a_1 \cdots a_n \in P$, any cut $H$ of $X$, and any vertex $v$ of $L_X$ (so $v \in \Z + \frac{1}{2}$ and $\frac{1}{2} \leq v \leq n+\frac{1}{2}$), 
define the \textbf{order} $\ord(H,v)$ to be the minimum number of chords (the edges connecting $i_\ell$ and $j_\ell$ if $H= \{(i_\ell, j_\ell)\}_{1 \leq \ell \leq m}$) of the
chord diagram $C_{X,H}$ that must
be crossed by any path from $v$ to the root which intersects the chord diagram
$C_{X,H}$ only transversely. If $v$ is the root, we define $C_{H,rt} := 0$.
\end{defn}
\begin{defn}
For any path $X$ and cut $H$, let the \textbf{order} of $H$, $\ord(H)$, be defined to be the maximum of all $\ord(H,v)$ for vertices $v$ of $L_X$.
\end{defn}
We note that a cut is simple iff its order is one.
\begin{defn}
  For two disjoint cuts $H_1, H_2$ of a path $X = a_1 \cdots a_n$, we
  say that $H_1 \prec H_2$ if, from all points, there is a path,
  intersecting $H_1$ and $H_2$ transversally, to the root that does
  not intersect chords from $H_1$ after those from $H_2$. In other
  words, the chords from $H_2$ are not separated from the root by any
  chords from $H_1$. 
\end{defn}
In particular, $\emptyset \prec H$ and $H \prec \emptyset$. Note that
$\prec$ satisfies the transitivity property. 

 We now may give the
formula
\begin{equation}
(\Delta \o 1) \Delta(X) = \Delta(X) \o 1 + \sum_{1 \leq \ord(H) \leq 2} 
\varepsilon_H \sum_{\text{simple cuts } H_1, H_2 \text{ such that } H = H_1 \sqcup H_2, H_1 \prec H_2} l_{H_1} \o l_{H_2,H_1} \o r_{H_2},
\end{equation}
where $l_{H_2, H_1}$ is the product of all paths along $L_X$ cut from the chord
diagram for $H$ which lie between a chord from $H_1$ and a chord from $H_2$.
 This implies the result.
\subsection{Proof of Theorems \ref{homt1}, \ref{homt2}} \label{homt12ps}
We prove Theorem \ref{homt2}, as well as injectivity of the
composition $D_* \circ S_*$, which implies Theorem \ref{homt1}.

First, we show that $S_*$ is a homomorphism of pre-Lie or Lie
coalgebras.  For this, we note that a cut $H$ together with a
specified chord $c \in H$ is the same information as a specified chord
$c$ together with two cuts $H_1, H_2$, one on each side of $c$, under
the correspondence $H = H_1 \sqcup H_2 \sqcup \{c\}$.  The datum
$(H,c)$ corresponds to a summand in the expression for
$\delta_* \circ S_*$ of a given (cyclic) path, while the datum
$(c,H_1,H_2)$ corresponds to a summand in the expression for
$(S_* \o S_*) \circ \delta_*$.  It is then easy to see that the two
summands are identical.

Next, we show that $S_{rt}$ extends to a homomorphism of Hopf algebras
\eqref{symchrtf}. We only need to check that $S_{rt}$ sends the
coproduct on $\Sym P$ to the coproduct on $\Sym Ch_{rt}$. For this, we
extend the observation of the previous paragraph: the datum $(H,H_1)$
of a cut $H$ and a simple subcut $H_1$ (corresponding to
a summand of $\Delta \circ S_{rt}(X)$) yields the same
information as the datum $(H_1, \{H_c'\}_{c \in H_1}, H_0')$, where $H_1$
is a simple cut, and $\{H_c', H_0\}_{c \in H_1}$ is a collection of cuts
on the connected components
$\{X_{H_1,c}\}_{c \in H_1} \sqcup \{X_{H_1,0}\}$ which result from
cutting $X$ along $H_1$ (corresponding to a summand of $(S_{rt} \o S_{rt}) \circ \Delta(X)$).  This correspondence is given by
$H = H_1 \sqcup \bigsqcup_{c \in H_1} H_c' \sqcup H_0'$.  Corresponding
data give identical summands of $\Delta \circ S_{rt}(X)$ and
$(S_{rt} \o S_{rt}) \circ \Delta$, given by 
\begin{equation}
\varepsilon_{H_1} l_{H_1}(X)_{\{H_c'\}_{c \in H_1}} \o r_{H_1}(X)_{H_0'},
\end{equation}
where the subscripts of $H_0', H_c'$ indicate which chord diagrams
to attach to the corresponding paths in the above monomial of paths.

Next, we show that $D_*$ is a homomorphism. First we tackle the pre-Lie, Lie
cases.  For any chord diagram $C_{X,H}$, we need to show that $(D_* \o D_*) \delta_*(C_{X,H}) = \delta_* \circ D_*(C_{X,H})$.  First, note that a choice
of chord of a chord diagram is the same as a choice of edge of the dual tree.
It is easy to see that the same tree-structure is obtained by either cutting
along this chord and then applying $D_*$ (dualizing), or applying $D_*$ first and then removing the corresponding edge.
It remains to show that, for any chord $c \in H$, we have
\begin{equation}
\varepsilon_{H \setminus c} \varepsilon_c = \varepsilon_H,
\end{equation}
which follows from the definition (and was first noticed in \cite{Tu2}).

In the Hopf algebra setting \eqref{chrtf}, showing $D_{rt}$ is a
homomorphism amounts to showing, for any chord diagram $C_{X,H}$, that
$(D_{rt} \o D_{rt}) \Delta(C_{X,H}) = \Delta (D_{rt}(C_{X,H}))$.
First, we note that a simple subcut of a chord diagram is the same as
a simple cut of the dual tree.  As before, it remains to show that the
signs work out correctly, that is, if $H_1$ is a simple subcut of $H$,
\begin{equation}
\varepsilon_{H_1} \varepsilon_{H \setminus H_1} = \varepsilon_{H}.
\end{equation}
This identity, noticed in \cite{Tu2}, is obvious from the definition.

Finally, we show that $D_* \circ S_*$ is injective.  This follows
because, for any (cyclic) path $X$, the trivial chord diagram yields a
summand of $D_* \circ S_*(X)$ which is the trivial tree (a point)
whose rooted or oriented tree-structure at that point includes $X$
itself.  All other summands are trees with $\geq 1$ edges.  So if we
compose $D_* \circ S_*$ with the projection to the space spanned by
the trivial tree with arbitrary structure, we easily obtain the
(cyclic) path $X$.

Note that, if the tree-structure is forgotten and we take the map to
the \cite{CK} algebra of trees itself, the composition is \textbf{not}
injective; e.g., any path without any pair of edges of the form
$e,e^*$ for $e \in Q$ must map to the trivial tree.

\footnotesize{
\bibliography{renorm-coalg}
\bibliographystyle{amsalpha}
{\bf W.L.G.}: Department of Mathematics, University of California, Riverside CA 92521, USA; \\
  \hphantom{x}\quad\, {\tt wlgan@math.ucr.edu} \\
{\bf T.S.}: Department of
  Mathematics, University of Chicago, 5734 S. University Ave, Chicago IL
  60637, USA;\\
  \hphantom{x}\quad\, {\tt trasched@math.uchicago.edu}
}

\end{document}